\documentstyle[12pt] {article}
\begin {document}
\title { Decoherence  and classical predictability of phase space histories.}
\author { C.\ Anastopoulos \\ Theoretical Physics Group, The Blackett
Laboratory \\
Imperial College, London SW7 2BZ, U.K. \\ E-mail : can@tp.ph.ic.ac.uk  \\  \\
PACS number = 03.65.Sq ; 05.40.+j}
 \date {June 1995}
\maketitle
 \begin {abstract}

\par
 We consider the decoherence of phase space histories in quantum
brownian
motion models,
consisting of a particle moving under a potential $V(x)$ in contact
with a
heat bath of temperature
$T$ and dissipation constant $\gamma$ in the Markovian regime. The
evolution of the
density operator for this system is thus
described by a non-unitary master equation. The phase space histories
are described by
quasiprojectors consisting of
gaussian density matrices smeared over large phase space cells. They
are characterized
by the size $[\Gamma]$ of the phase
space cell together with the size $[M]$ of the margin (the region at
the boundary
of $\Gamma$ in which the Weyl symbol of the
projector goes from $1$ to $0$). By generalizing an earlier result of
Hagedorn on the
unitary evolution of coherent states, we
show that an initial Gaussian density matrix remains approximately
Gaussian under
non-unitary time evolution, and derive a
bound giving the validity of this approximation. This result is then
used, following earlier work of Omn\`es to compute the time
evolution of the phase space projectors under the master equation:
The evolution of an initial projector $P$ onto a cell
$\Gamma$ is approximately equal to another phase space projector $P'$
onto a
cell $\Gamma'$, where $\Gamma'$ is the
classical dissipative evolution of $\Gamma$. Furthermore, the expected
unpredictability
due to thermal fluctuations is reflected
in the fact that the margin of $P'$ (and hence the effective region it
occupies
in phase space) is greater than that of $P$. We
thus show that histories of phase space samplings approximately
decohere, and
that the probabilities for these histories are
peaked about classical dissipative evolution, but with an element of
unpredictability
due to the noise produced by the
environment.
\end {abstract}
\pagebreak

\let \ssection = \section
\renewcommand{\section}{\setcounter{equation}{0} \ssection}

\section {Introduction}
\par
Ever since the first days of quantum mechanics, the question of the
emergence of the classical
deterministic  world
from the underlyinfg probabilistic laws is considered of particular
importance. It is closely
connected to the measurement
problem, since any measurement involves the interaction of the system
under observation with a
large deterministic device.
\par
Different aspects of this problem have been adressed extensively in
the literature.
Still, there are issues that remain unclear.
Among them are question of the validity of the description of many
body systems with
collective variables that evolve under
quasiclassical equations of motion, the conditions under which
classical predictability
arises in a quantum system, its possible breakdown at
long times and how it is affected by environmentally induced noise. In
this paper we will
address these problems, as appearing
in a wide class of open quantum systems, namely quantum Brownian motion models.
\par
Our analysis lies within the framework of the decoherent histories
approach to
quantum mechanics, as set out by
 Griffiths \cite{Gri},Omn\`es \cite{Omn1,Omn2,Omn3} and Gell-Mann and
Hartle \cite{GeHa}.
The main element of this formalism  is the notion of history.
 A history is a string of projection operators at consecutive moments
of time. We can
build sets of histories, by taking a
partition  of the unit operators into projectors $P_{\alpha}(t)$ at
each moment of time.
These operators satisfy:
\begin{eqnarray}
P_{\alpha} P_{\beta} = \delta_{\alpha \beta} P_{\alpha} \\
\sum_{\alpha} P_{\alpha} = 1
\end{eqnarray}
We usually consider them to evolve in the Heisenberg picture
\begin{equation}
P_{\alpha}(t) = e^{iHt} P_{\alpha} e^{-iHt}
\end{equation}
An element
 of a set of such histories reads :
\begin{equation}
C_{\alpha} = P^n_{\alpha_n}(t_n) \ldots P^2_{\alpha_2}(t_2) P^1_{\alpha_1}(t_1)
\end{equation}
\par
A space of histories has a natural candidate probability measure
\begin{equation}
p(C_{\alpha}) = Tr (C_{\alpha} \rho_0 C_{\alpha} )
\end{equation}
This measure can be interpreted as a probability only if the standard
probability sum rules are satisfied.
 The set of histories is then called consistent (or decoherent). The
condition for this
can be writen in terms of
the decoherence functional , a complex valued function of pairs of histories:
\begin{equation}
D(\alpha,\alpha') = Tr(C_{\alpha} \rho_0 C_{\alpha'})
\end{equation}
The necessary and sufficient condition for the sum rules to be satisfied is
then
\begin{equation}
Re D(\alpha,\alpha') = 0      \qquad \alpha \neq \alpha'
\end{equation}
We usually employ the stricter condition
\begin{equation}
 D(\alpha,\alpha') = 0      \qquad \alpha \neq \alpha'
\end{equation}
The terms consistency and decoherence of the set of histories are
given in the
literature to the conditions (1.7) and (1.8)
respectively. That assignment of a probability measure to  a set of
histories,
allows us to reason with the histories using the
rules of classical logic.
\par
In general, decoherence appears in physical systems, only for coarse grained
histories.
A typical
example of coarse graining is to ignore a number of degrees of freedom
of the system,
or take projectors into large intervals
of position or momentum. The decoherence is in almost all physical
situations only
approximate, that is, the probability sum
rules are satisfied only within an order of $\epsilon$, where
$\epsilon$ is a small
positive number.
\par
In quantum mechanics the most general proposition for a physical
system at a given
instant of time is represented by
 a projection operator. In classical mechanics the corresponding
entity is the phase
space cell. Thus, when we want
 to discuss classical deterministic behaviour, we have to construct
entities that
bridge the gap between the
 two formalisms.
\par
This is done through the introduction of quasiprojectors, that is,
approximate
projectors onto phase space cells.
 They are constructed in such a way, that their corresponding symbol
in the
Wigner representation is a smoothened
characteristic function onto some phase space cell. Omn\`es
\cite{Omn2} has
shown that histories corresponding to
classical hamiltonian evolution of sufficiently large and regular
phase
space cells, approximately satisfy the
 consistency conditions. Also the conditional probability for our
system to
be found in the cell $\Gamma_2$ at
time $t_2$, if at time $t_1$ it was within the cell $\Gamma_1$, is
almost unity,
provided $\Gamma_2$ is obtained from
$\Gamma_1$ through evolution according to the classical equations of motion.
\par
In this paper, we are interested in generalizing these results for a class of
open
systems, namely quantum Brownian motion models. These consist of a
particle
moving under a potential
 $V(x)$ and coupled to a large system which is taken as the
environment.
Usually we take as an environment a
collection of harmonic oscillators in a thermal state. By tracing out
the
environment degrees of freedom, we
 obtain an effective evolution equation for the density matrix of the
particle.
The coupling of the environment to
the particle is contained within two effects: dissipation and diffusion.
\par
We concentrate on a particular case of those models, namely when the evolution
equations are Markovian.
 This corresponds to the case of ohmic environment at high temperature.
This is mainly due to the fact that only in the Markovian limit can we
write
the decoherence functional in terms of the
 reduced
density matrix.
\par
These models can be used as prototypes for more complicated systems,
as for instance  particle detectors in early
universe or continuous measurement. What is more interesting for our
approach, is that we can model this way the
effective behaviour of collective or hydrodynamic variables. We can
identify the degrees of freedom of the particle with distinguished
variables in a many-body system, like for instance the center of mass
position and momentum of a macroscopic body,
while the environment represents all the ignored degrees of freedom
like
the positions and momenta of its constituents.
\par
The issue we are concentrating on, is the conditions under which the
description of the system with the classical
dissipative equations of motion can be accurate. Essentially we want
to prove
approximate determinism for quantum
dissipative systems. Our approach is rather different from other
authors
that have discussed decoherence of
 phase space histories \cite{Bru,Twa} or configuration space histories

\cite{DoHa} and closer to the spirit
of Omn\`es \cite{Omn2,Omn3} .
\par
Our main results can be summarized, as follows:
We establish the degree of the validity of the gaussian approximation
for the
Markovian regime of quantum
 Brownian motion models. We show that a gaussian density matrix
remains
approximately gaussian centered
 around the classical path for a large class of potentials $V(x)$.
The validity of the approximation is quite good
as long as the spread in position is much smaller than the scale on
which the non-linearities in the potential
 become significant.
\par
We use this result to show, that quasiprojectors onto sufficiently
large and
smooth phase space cells evolve
according to the classical equations of motion. The environment
induces a degradation on the evolved projectors,
 which leads essentially to a loss of predictability.
\par
Finally, we construct the decoherence functional in the Markovian
regime and
establish that histories corresponding
to classical evolution on phase space approximately satisfy the
consistency
condition, and thus  approximate
 determinism arises.
\par
This paper is structured as follows:
After giving  a brief review of quantum Brownian motion models, we
proceed to
discuss phase space projectors in
 section 3. We generalize the construction of quasiprojectors from
coherent
states of Omn\`es, using gaussian
 density matrices. In section 4 we state a theorem on the degree of
validity of the gaussian approximation in
our models, using which we establish how the quasiprojectors we
constructed evolve in Section 5. We construct
 histories corresponding to classical evolution in Section 6 and
discuss their decoherence properties. In the last
 section we summarize and discuss our results.
\par
The notation we will use is as follows:
 Whenever there is a possibility of confusion the reduced density
matrix
will be denoted as $\tilde{\rho}$,
 otherwise plainly as $\rho$.
The Hilbert space of the wave functions will be denoted as $H_{\psi}$
and the one of the density matrices under
the Hilbert-Schmidt inner product as $H_{\rho}$. We write the inner
product
in $H_{\psi}$ using the Dirac notation
 $\langle \psi_1 |\psi_2 \rangle$  and in $H_{\rho}$ as
$(\rho_1,\rho_2)$ .
The action of an operator
${\cal L}$ on an element of $H_{\rho}$
 will be denoted as
${\cal L}[\rho] $.
The volume of a phase space cell $\Gamma$ is written $[\Gamma]$. In
Section 4
and in the Appendix we will use
units such that $\hbar$ is dimensionless.

\section { Quantum Brownian Motion Models }

\par
We first give the basic features of the model, within  which we are
going
to discuss
phase space decoherence.  We consider a particle of mass $M$ moving under
 the influence of a potential $V(x)$ and in contact with a  heat bath
of
harmonic oscillators
\cite{CaLe,HPZ,GSI}.
\par
The total action of the system reads :
\begin{eqnarray}
 S_{tot} [x(t),q_n(t)] = \int dt [\frac{1}{2} M \dot{x}^2 - V(x)] +
\sum_n  \int dt  [\frac{1}{2} m_n
\dot{q_n}^2   \nonumber \\
- \frac{1}{2} m_n \omega_n^2 q_n^2 - C_n q_n x ]
\end{eqnarray}
We concentrate in the behaviour of the distinguished particle and
thus construct
the reduced density matrix:
$$  \tilde{\rho}_t(x,x') = \prod_n \int dq_n \rho_t(x,q_n,x',q_n) $$
We can describe its time evolution using the reduced density matrix
propagator $J$,
defined by the relation:
$$ \tilde{\rho}_t (x,y) = \int dx_0 dy_0 \quad J(x,y,t| x_0,y_0,0)
\tilde{\rho}(x_0,y_0)  $$
 Under the assumption that the initial density matrix $\rho_0$
factorizes, the propagator is given by
the path integral expression :
\begin{equation}
 J(x_f,y_f,t \mid x_0,y_0,0) = \int Dx Dy \exp ( \frac{i}{\hbar} S[x] -
\frac{i}{\hbar} S[y] +
\frac{i}{\hbar} W[x,y])
\end{equation}
where
$$ S[x] = \int_{0}^{t} dt  (\frac{1}{2} M \dot{x}^2 - V(x) ) $$
$W[x(t),y(t)]$ is the Feynman Vernon influence functional phase
\begin{eqnarray}
W[x(t),y(t)] = - \int_{0}^{t} ds \int_{0}^{s} ds' [x(s)-y(s)]
\eta(s-s')
[x(s') + y(s')] \nonumber \\
+ i  \int_{0}^{t} ds \int_{0}^{s} ds' [x(s)-y(s)] \nu (s-s') [x(s') - y(s')]
\end{eqnarray}
The path integration is taken over all paths satisfying $x(0)
= x_0$, $x(t) = x_f$.
The kernels $\eta(s)$ and $\nu(s)$ are defined by :
\begin{equation}
 \nu(s) = \int_0^{\infty} \frac{d \omega}{\pi} I(\omega)
\coth(\frac{\hbar \omega}{2 k T }) \cos \omega s
\end{equation}

\begin{equation}
 \eta(s) = \frac{d}{ds} \gamma(s)
\end{equation}
where
\begin{equation}
 \gamma(s) = \int_0^{\infty} \frac{d \omega}{\pi}
\frac{I (\omega)}{\omega} \cos \omega s
\end{equation}
and $I(\omega)$ is the spectral density,
\begin{equation}
 I(\omega) = \sum_n \delta(\omega - \omega_n) \frac{\pi C_n^2}{2 m_n \omega_n}
\end{equation}

\par
The kernel $\nu(s)$ modifies the action of the distinguished system
and leads to dissipation and
renormalization of the potential. The kernel $\eta (s) $ is
responsible for noise and the process of decoherence.
 Both kernels are completely determined once a form for the
spectral density is
specified. A choice that proves to be convenient is
\begin{equation}
 I(\omega) = M \gamma \omega (\frac{\omega}{\Lambda})^{s-1}
exp (-\frac{\omega^2}{\Lambda^2})
\end{equation}
where $\Lambda$ is a cut-off, which is usually
taken to be very large.
\par
The ohmic case ($s=1$) is of particular interest.
Taking $\Lambda$ to infinity gives
$$ \gamma(s) = M \gamma \delta (s)  $$
which means that the corresponding classical equations of motion are
local in time. The noise
kernel is non-local for large $\Lambda$, except in the Fokker-Planck
limit, $ kT >> \hbar \Lambda $,
in which we have,
\begin{equation}
 \nu (s) = \frac{2 M \gamma k T}{\hbar} \delta (s)
\end{equation}

 In this regime the reduced density matrix $\rho$ (we drop the tilde)
satisfies the master equation
\begin{eqnarray}
 \frac{\partial \rho}{\partial t} = &-& \frac{\hbar}{2 M i}
( \frac{\partial^2 \rho}{\partial x^2} - \frac{\partial^2 \rho}{\partial y^2} )
 - \gamma (x-y) ( \frac{\partial \rho}{\partial x}
- \frac{\partial \rho}{\partial y} ) \nonumber \\
 &+& \frac{1}{i \hbar}
 [ V(x) - V(y) ] \rho - \frac{D}{\hbar^2} (x - y)^2 \rho
\end{eqnarray}
where we set  $ D = 2 M \gamma k T$,
or in operator form :
\begin{eqnarray}
 \frac{\partial \rho}{\partial t} &=& \frac{1}{i \hbar}
[ \frac{p^2}{2 M} + V(x) , \rho ]
\nonumber \\
&-&\frac{\gamma}{ i \hbar} [x,\{ \rho,p\}] -
\frac{D}{\hbar^2} [x,[x,\rho]]   \nonumber \\
&\equiv& {\cal L}[\rho]
\end{eqnarray}
\noindent This master equation describes a Markovian process. The time
evolution
of the initial density matrix can be
 represented by
 the action of a one-parameter semigroup with generator ${\cal L}$
on the state space.

We will represent an element of the semigroup as $ e^{{\cal L}t} $
and the time evolution of an initial density matrix
 $\rho_0$ as $e^{{\cal L}t}[\rho_0]$ .

\pagebreak

\section {Phase space projectors}
\subsection{General considerations}
\par
In order to discuss phase space histories, we have to find elements
of the quantum mechanical formalism
that correspond closely to the notion of a phase space cell,
at least in the macroscopic domain.
 Since any proposition about a physical system at a given
instant of time
can be represented as a projection operator, what we need is
a class of operators that project onto
 the subspace of physical states which  have position and momentum
well within some phase space cell $\Gamma$.
\par
Position and momentum cannot be simultaneously specified with
arbitrary
accuracy, so we must consider approximate
projection operators . We call an operator $P$ a
quasiprojector into a phase space cell $\Gamma$ if any wave function
well localized inside $\Gamma$ (outside $\Gamma$ )
 is an approximate eigenfunction of $P$ with eigenvalue 1 ( 0 ) .
The construction of such operators is particularly easy in
the Weyl representation, in which we associate with $P$ a function
$f(x,\xi)$ on a mock phase space, which we call its associated symbol
\begin{equation}
 f(x,\xi) = \int dy \exp (-i \xi y/ \hbar) \langle x
+ \frac{y}{2} | P | x - \frac{y}{2} \rangle
\end{equation}
\par
In general, a function $f(x,\xi)$ on a two dimensional phase space is
called
a symbol of order $m$, where
 $m$ is real, if its derivatives of all orders have the following bounds
\begin{equation}
 | \partial^{\alpha}_{x} \partial^{\beta}_{\xi} f(x,\xi) | \leq
C_{\alpha \beta}
( 1 + x^2 + \xi^2 )^{(m- \alpha - \beta)/2}
\end{equation}
\noindent where $C_{\alpha \beta}$ are constants called the seminorms
of $f$ .
Note also, that $x$
 and $\xi$ are in dimensionless units (scaled by some typical
dimensions $L$ and $P$ ).
   \par
One possibility is to take for the symbol corresponding to a
quasi-projector
on a phase space cell
$\Gamma$ the characteristic function of the cell $\Gamma$.
This does not define a good operator because of the
 discontinuity at the boundary so instead, we consider a smoothened
characteristic function \cite{Omn3}.
That is, we take $f$ to be equal to 1 inside and to 0 outside
$\Gamma$,
except in a small region along
the boundary of $\Gamma$, where it interpolates smoothly from 1 to 0.
This transition region is called
the margin $M$ of the cell $\Gamma$ . In order for the operator to be
well defined, the phase space cell has to
be regular; loosely speaking it should not develope structure on the
scale
of $\hbar$ and the volume
 of the margin $M$ should be much smaller than the volume of $\Gamma$. The
parameter
 $\epsilon = [M]/[\Gamma] $
is a measure of the validity of the approximation. For a sufficiently
regular cell it is of the order of
$ (\hbar/LP)^{\frac{1}{2}}$, where $L$, $P$ are
typical length and momentum scales for our cell. We will return to the

issue of regularity in more detail later.
The quasiprojectors $P$ have a number of interesting properties
\cite{Omn2}:
\\

 {\bf 1.} Their trace is proportional to the volume of the phase-space cell.
\begin{equation}
  TrP = [\Gamma]/(2 \pi \hbar)
\end{equation}

 {\bf 2.} They are very close to true projectors:
\begin{equation}
  Tr|P-P^2| < c \epsilon
\end{equation}
where $c$ is a number of the order of unity.

 {\bf 3.} For two different projectors on the same phase space cell
$P$
and $P'$  with
corresponding parameters $\epsilon$ and $\epsilon '$ we have :

\begin{equation}
         \qquad Tr|P-P'| < c (\epsilon + \epsilon ' )
\end{equation}

 {\bf 4.}  If $\Gamma_1$, $\Gamma_2$ and $\Gamma_1 \cap \Gamma_2$ are
three phase space cells and $P_1$, $P_2$, $P_{12}$
quasiprojectors
 associated to them, the operator
$ \delta P = P_1 P_2 - P_{12} $ is bounded in trace norm by :
\begin{equation}
 Tr|\delta P| < K \frac{[\Gamma]}{2 \pi \hbar} (\hbar /LP)^{\alpha}
\end{equation}
where $[\Gamma] = sup([\Gamma_1],[\Gamma_2])$ , $\alpha > \frac{1}{2}
$
and $K$ of order unity.
\subsection{Coherent state quasiprojectors}
\par
One particular way to construct quasiprojectors is through the use of coherent
states \cite{Omn2}. Consider the gaussian wave function :
\begin{equation}
 g_{qp} (x) = (\frac{\Sigma}{2 \pi \hbar})^{\frac{1}{4}}
\exp ( -\frac{\Sigma}{4 \hbar} (1 + i r) (x-q)^2 +
i p x/ \hbar )
\end{equation}
\noindent This induces a metric :
\begin{equation}
 d(x,p) = \frac{\Sigma}{4 \hbar} x^2 + \frac{\hbar}{\Sigma (1 + r^2)} p^2
\end{equation}
on phase space with respect to which one can give a precise definition
of
a regular phase space cell \cite{Omn2,Omn3}.
\par
We call a phase space cell regular to order $\epsilon$, with respect
to
one particular family of coherent states if:
\\

 {\bf 1.}  The curvature radii with respect to the metric $d$
of $\partial \Gamma$ are larger than $l$ in absolute
value.

 {\bf 2.}  The margin of the cell $M$ which is defined as :
$$ M = \bigcap_{(x,p)\epsilon \partial \Gamma}  e(x,p,l) $$
where $e(x,p,l)$ is the ellipsis  defined by :
$$e(x,p,l) = \{ (x',p') | d(x-x',p-p') < l^2 \}  $$
satisfies  $[M] < \epsilon [\Gamma]$ .

 {\bf 3.} The numbers $\epsilon$ and l satisfy : $ e^{-2 l^2} < \epsilon $.
\\
\par
If $\Gamma$ is a regular cell to order $\epsilon$ then the operator :
\begin{equation}
 P = \int_{\Gamma} \frac{dq dp}{2 \pi \hbar} |g_{qp} \rangle \langle g_{qp}|
\end{equation}
is a quasiprojector associated with $\Gamma$. We can readily see that its
symbol :
\begin{eqnarray}
 f(x,\xi) =  \int_{\Gamma} \frac{dq dp}{ \pi \hbar} \exp [ -
\frac{\Sigma}{2 \hbar}
(1 + r^2) (x-q)^2 \nonumber \\
 - \frac{2}{\hbar \Sigma} (\xi - p)^2 - \frac{ 2 \Sigma r}{\hbar} (x-q)(\xi-p)
]
\end{eqnarray}
is up to corrections of order $e^{-l^2}$  equal to 1 (0) inside
(respectively outside) $\Gamma$ except
for the margin M where it interpolates between those two values.
It also satisfies the conditions for being
a symbol of arbitrary negative order.
\subsection{Generalized gaussian projectors}
\par
 We are mainly interested in the time evolution of the
quasiprojectors,
when our system is under
the influence of the thermal environment, as in the case of quantum
Brownian motion models.
Due to the non-unitarity of the evolution, the operator in time $t$
will
take a form that cannot be  related to the
 quasiprojectors (3.9) in a straightforward way, as in \cite{Omn2}.
Omn\`es has shown that under a large class of potentials, the density operators
of the form $ |g_{qp} \rangle \langle g_{qp} | $ evolve
approximately into operators of the same form but belonging to a different
family of coherent states (different parameters $\Sigma$ and $r$).
The non - unitarity of the evolution in
our case implies that an operator of this form become mixed.
Therefore, we need a larger class of gaussian density operators.
\par
 We propose a class of quasiprojectors, that are defined through
gaussian density matrices,
which we choose to parametrize by the set of five numbers $(\Sigma,F,r,q,p)$ as
\begin{eqnarray}
 \langle x|\rho|y \rangle = (\frac{\Sigma}{2 \pi \hbar})^{\frac{1}{2}}
\exp [ - \frac{\Sigma}{2 \hbar} (\frac{x+y}{2} - q )^2
\nonumber \\
- \frac{F}{2 \hbar} (x-y)^2 - \frac{i r \Sigma}{2 \hbar}
(\frac{x+y}{2} - q) (x-y)
+ i \frac{p}{\hbar} (x - y) ]
 \end{eqnarray}
which is defined for $ \Sigma \leq 4 F $ in order to satisfy the positivity
requirement.
For these density matrices we can verify that,

\begin{equation}
 \|\rho\|_{H.S.} = (Tr \rho^2 )^{\frac{1}{2}} = (\Sigma/ 4
F)^{\frac{1}{2}} =
(\hbar /2 {\cal A}) \leq 1
\end{equation}
where ${\cal A}$ is the Wigner function area defined as :
$$ {\cal A}^2 = (\Delta q)^2 (\Delta p)^2 - C_{pq}^2  $$
with :
$$C_{pq} = \frac{1}{2} \langle (q-\langle q \rangle)(p-\langle p
\rangle )+
(p-\langle p \rangle)(q- \langle q \rangle )
   \rangle   $$
${\cal A}$ is a measure of the phase space area in which the gaussian
density matrix is localized.
\par
This density matrix defines a metric on phase space in analogy to (3.8):
\begin{equation}
 d(x,p) = \frac{\Sigma}{4 \hbar} x^2 + \frac{\hbar}{4 F(1 +
\frac{\Sigma}
{4 F} r^2)} p^2
\end{equation}

With respect to this metric, we can define the notion of the margin of
a phase space cell and the notion of
 regularity (up to order $ \epsilon$) as before .
We define the operator :
\begin{equation}
 P = \int_{\Gamma} \frac{dq dp}{2 \pi \hbar}  \quad \rho(\Sigma,F,r,q,p)
\end{equation}
This is a quasiprojector associated with the cell $\Gamma$.
\par
To see this, consider its symbol,
\begin{eqnarray}
 f(x,\xi) = (\Sigma/4 F)^{\frac{1}{2}}  \int_{\Gamma} \frac{dq dp}{
\pi \hbar}
\exp [-\frac{\Sigma}{2 \hbar} (1 + \frac{\Sigma}{4 F} r^2)
(x - q)^2  \nonumber \\
- \frac{1}{2 \hbar F} (\xi - p)^2 -\frac{\Sigma r}{2 \hbar F} (x - q)(\xi - p)
]
\end{eqnarray}
It is a symbol of arbitrary negative order and has the form of the
most general
gaussian smeared characteristic
 function associated to the cell $\Gamma$. We can verify that it takes
values 1 ( 0 )
inside (outside) $\Gamma$ up to
 corrections
 of order $e^{-l^2}$, where $l$ is defined as before but with respect
to the
metric (3.13). It interpolates between
 those two values only in the margin, which has a width of order ${\cal
A}^{\frac{1}{2}}$ .
\par
How can we compare this class of projectors to the ones defined through the
general gaussian states (3.11)? A coherent
state is localized within a volume of order $\hbar/2$ and is unable to
distinguish
points found within a volume of this size. For this reason we expect the margin
associated with a projector onto a phase
 space cell to have an area of order $ \hbar$ and the parameter
$\epsilon$
to be of order
$(\hbar /LP)^{\frac{1}{2}}$. This estimation
is backed by a detailed calculation using microlocal analysis, that
is, we
can verify that this order of magnitude
 for the parameter $\epsilon$ is the one that gives the best
approximation
of a quasiprojector to an exact
 projector \cite{Omn2,Omn3}.
\par
On the other hand the gaussian density matrix  (3.11) is localized within a
volume
of order ${\cal A}$. Therefore we expect the width of the margin to be
of order
${\cal A} $ and the parameter $\epsilon$ to be of order
 $({\cal A}/LP)^{\frac{1}{2}}$ for the class of projectors
(3.14). This is
clearly, not the ``best" quasiprojector one can
 associate to a phase space cell, but this generalization is
necessary. As long as the quantity $ (2 {\cal A}/
\hbar)^{\frac{1}{2}}$ is of
the order of
unity we still have a good approximation for a phase space projector.
The larger the value of ${\cal A}$ is,
the worse the degree of approximation to a true projector is. We are
therefore going to refer to projectors with
 minimum value of ${\cal A} = \hbar/2$ (the ones constructed from pure
gaussians)
as the maximum resolution projectors.

\par
We will see that the time evolution of an initial generalized coherent
state
remains within a
good approximation a gaussian density matrix, with time increasing
value of ${\cal A}$.
We know from the study
 of linear systems that the value of ${\cal A}$ increases polynomially
in the short time limit,
to become constant for times much larger than the relaxation time
$\gamma^{-1}$ \cite{AnHa}. Its asymptotic value
 corresponds to the uncertainty due to thermal fluctuations, and is
typically many orders
 of magnitude larger than the quantum ones. Still, for sufficiently large phase
space cells ${\cal A}$ is much smaller than the size of the cell, the quantity
 $\epsilon$ remains quite small and our operators (3.14) are still
close
to true projectors. On the other
 hand, for smaller cells for which $ \hbar << LP < {\cal A}_{\infty} $ there
comes some time (usually of the order
 of $\gamma^{-1}$), when the evolved quasiprojectors cannot anymore distinguish
the phase space cell,
 since the size of the margin has become essentially as large as the whole of
the cell.

\section { Time evolution of gaussian density matrices }
  \par
We are interested in the evolution of the gaussian density matrices
(3.11)
in the class of quantum
Brownian motion models described by the master equation (2.11). As we
mentioned,
 this expression is valid in the Markovian regime (high temperature
and
ohmic environment).
\par
For linear potentials, we know (see for instance \cite{CaLe,AnHa} )
that the
propagation is gaussian,
 and an initial gaussian density matrix remains gaussian centered
around the
classical path. Also in the case of zero
 coupling to the environment (unitary evolution) Hagedorn \cite{Hag}
has
established that a generalized
coherent state retains its shape for a period of time  and its center
follows the classical equations of motion. The error
 of this approximation is of the order of $(\hbar / LP ) ^{\lambda}$,
where $\lambda < \frac{1}{2} $.
\par
We seek a generalization of these results. We want to show, that for a large
class of
potentials $V(x)$ the master equation (2.11) respects the gaussian nature of
the
density matrices (3.11) and
that their centers follow the classical equations of motion (with
dissipation).
 We shall show that this is a good
 approximation as long as the spread in position is sufficiently small
and
that the error is of the order
 of $(\hbar/LP)^{\lambda} $ as in the purely hamiltonian case.
\subsection{The theorem}
\par
 In this section, we  present our result in the form of a theorem, a
detailed proof of which can be found in the Appendix.
   We assume that the potential $V(x)$ satisfies:
\\

 {\bf1.}   $V^{(2)}(x)$ is continuous and uniformly Lipschitz on
compact
subsets of ${\cal R}$ (i.e. given any $R >0$
 there exists $k$ such that $|V^{(2)}(x) - V^{(2)}(y) | < \beta |x-y|
$
whenever $|x| < R $,
$|y| < R$.

 {\bf 2.}  $|V(x)| < e^{M x^2} $

 {\bf 3.}  $V(x)$ is bounded from below.
\\
\par
Assume that at $t = 0$ the system is in the state
$\rho_0 = \rho(\Sigma_0,F_0,r_0,q_0,p_0)$. The exact solution
 to the master equation (2.11) can be writen formally as :
$$ \rho_t = e^{{\cal L}t} [\rho_0]$$
where ${\cal L}$ is the generator of the one-parameter semigroup
acting
on the state space, which is determined by the
dynamics.
\par
Then we have the following theorem :
\par
\indent {\bf Theorem:}
For each $T > 0$, $N > 0$, $p < \frac{1}{2}$, $ \alpha < \frac{1}{2} -
p $
and $ 0 < t < T $, such that
$$(\frac{\Sigma F }{\Sigma + 4 F})^{-1} < N \hbar^{-2p}  $$
 there exists $C > 0$, $\delta > 0$ for which :
\begin{equation}
 \| e^{{\cal L}t} [\rho_0] - \rho(\Sigma(t),F(t),r(t),q(t),p(t))
\|_{HS}
< C \hbar^{\lambda}
\end{equation}
whenever $\hbar < \delta$. We denote $ \lambda = 3 \alpha -1 $.
\par
The quantities $(\Sigma(t),F(t),r(t),q(t),p(t))$ are solutions to the
system of differential equations :
\begin{eqnarray}
 \dot{q} &=& \frac{p}{M}  \\
 \dot{p} &=& -V'(q) - 2 \gamma p  \\
 \dot{\Sigma} &=& \frac{1}{M} \Sigma^2 r  \\
 \dot{F} &=& \frac{1}{M} \Sigma F r - 4 \gamma F + \frac{2D}{\hbar} \\
 \dot{r} &=& - \frac{\Sigma r^2}{2 M} - \frac{2}{M} F - 2 \gamma r
+ \frac{2}{\Sigma} V^{(2)}(q)
\end{eqnarray}
under the initial conditions :
\begin{equation}
 ( \Sigma(0),F(0),r(0),q(0),p(0) ) = ( \Sigma_0,F_0,r_0,q_0,p_0)
\end{equation}
Note that we have used dimensionless units with $L=1$ and
$P=1$.
\par
The proof is based on the observation that the generator ${\cal L}$ can be
split
into two parts, described in the Appendix:
$ {\cal L} = {\cal S} + {\cal U} $, where  $ {\cal S}$ is the
generator that
corresponds to the Brownian motion of a free particle
and ${\cal U}$ a generator that contains only the effects of the potential. The
action of
${\cal S}$ can be exactly
determined (for instance using the density matrix propagator which can
be exactly
computed for the free particle),
while the action of ${\cal U}$ can be approximated
  with a quadratic expression. Thus we get expressions for
$e^{{\cal S} t} [\rho_0]$ and
$e^{{\cal U}t} [\rho_0]$ which
 can be combined to give an expression for $e^{{\cal L} t} [\rho_0]$
through the use of the Trotter
product formula. The proof follows closely the one of Hagedorn \cite{Hag}, with
the main differences being the non-unitarity of the generator ${\cal S}$ and
that our
Hilbert space is the one of the density matrix under the
Hilbert-Schmidt norm.
\par
The proof involves taking a Taylor expansion of the potential and
keeping
only the quadratic terms.
 Therefore our results are exact for the harmonic oscillator and the
free
particle case.
\subsection{ Uncertainties}
\par
We can write the uncertainties in position $\Delta q$ and in momentum
$ \Delta p$ as well as the correlation
$ C_{pq}$ associated with the gaussian density matrix in terms of the
parameters $\Sigma$, $r$ and $F$:
\begin{eqnarray}
 (\Delta q)^2 &=& \hbar/\Sigma   \\
 (\Delta p)^2 &=& \hbar F ( 1 + \frac{\Sigma}{4 F} r^2)    \\
 C_{pq} &=& -\frac{\hbar r}{2}
\end{eqnarray}
{}From this we can see the interpretation of the parameters $\Sigma$,
$F$
and $r$ we used to parametrize the
gaussian density matrices. Clearly $\Sigma^{-1}$ is a measure of the spread in
position and $r$ of the correlation between position and momentum.
$\Sigma/F$ is proportional to $Tr \rho^2$ and therefore is a measure
of the non-purity of the state. The original parameters
$\Sigma$, $F$, $r$ have been more convenient in the course of the
proof,
but in order to have a clearer interpretation
 of the results, we shift our attention to the set of variables
$\Delta q$,
$\Delta p$, $C_{pq}$.
Within our approximation they evolve according to the equations:
\begin{equation}
 \frac{d}{dt} (\Delta q)^2 = \frac{2}{M} C_{pq}
\end{equation}

\begin{equation}
 \frac{d}{dt} (\Delta p)^2 = - 4 \gamma (\Delta p)^2
- 2 C_{pq} V^{(2)}(q) + 2 D
\end{equation}

\begin{equation}
 \frac{d}{dt} C_{pq} = \frac{1}{M} (\Delta p)^2 -2 \gamma C_{pq}
- (\Delta q)^2 V^{(2)}(q)
\end{equation}

\par
We notice, that the diffusion coefficient appears only in the equation

for the momentum uncertainty, and at
short times is the dominant term. This means that the spread in
momentum
is more effective than the corresponding
spread in position, which depends on the diffusion coefficient only
indirectly.
It is also important to stress the reluctance of
particles with large mass to undergo a rapid growth of the fluctuations.

\subsection { Validity of the approximation }
\par
The essential requirement for the validity of our approximation is
that the
quantity $\frac{\Sigma +4 F}{\Sigma F} $
 remains sufficiently large through the evolution according to the
equations
(4.2-4.6). Since we have that
$ \Sigma \leq 4 F $, this is equivalent to saying that $ \Sigma $
remains
sufficiently large
(much larger  than  $N \hbar^{2p}$). But a large value of $\Sigma$ means that
 the particle is well localized in position. We can thus say, that while the
state remains well localized
in position (that means $\Delta q < N^{-\frac{1}{2}}\hbar^{\frac{1}{2} - p}$),
the gaussian approximation will be quite good. When, mainly due to the
diffusion,
the state of the particle has become spread in space, the large scale
structure
of the potential becomes important and even weak non-linearities will
contribute significantly
in the evolution, thus rendering the gaussian approximation invalid. Note, that
the
constants $\lambda$ and $p$ satisfy :
$$ \lambda < \frac{1}{2} - 3 p$$
which shows that the larger our tolerance for the spread in position,
the larger
the error stemming from our
 approximation.
\par
In general, we cannot say much about the time when the gaussian
approximation
breaks down, since this depends
 crucially on the form of the potential, the particle mass and the
position and spread of the initial gaussian.

 Excluding cases of potentials varying significantly in the
microscopic scale or
potentials where tunnelling
effects can be important, we expect the gaussian approximation to hold
within a
very good accuracy for times
 much smaller than the typical time scale in our dynamics :
$\gamma^{-1}$.
For larger times, the validity
 of the approximation, depends heavily on the length scale $L$ on
which
$V^{(2)}(x)$ varies. If this length scale
 is much larger than the size of the thermal fluctuations the
approximation
will hold for timescales some orders
 of magnitude larger than $\gamma^{-1}$. If this is not the case,
the approximation
will have broken down much earlier.
Within a few times $\gamma^{-1}$ the particle will tend towards the
state
of thermal equilibrium.
 \par
We can clarify those ideas by examining the simplest case of a system
moving
in a potential exhibiting only
 weak nonlinearities :
\begin{equation}
 V(x) = \frac{1}{2} M \omega^2 x^2 + \eta x^4
\end{equation}

This effect of the non-linearities becomes significant at length
scales of
the order of
$$ L = (\frac{M \omega^2}{\eta})^{\frac{1}{2}}$$
 Assuming weak coupling to the environment, the uncertainty in
position
is given in leading order to
 $\epsilon$ and $\gamma/ \omega$ \cite{AnHa} :
\begin{equation}
 (\Delta q)^2 = (\Delta q')^2 e^{-2 \gamma t} + \frac{M k T}{\omega ^2} ( 1 -
e^{-2 \gamma t})
\end{equation}
with $(\Delta q')^2$ containing the effects of the unitary evolution.
The gaussian approximation will break down when $ \Delta q \simeq
L$. We can readily verify that for
 $ \eta <<  \omega^{4}/k T   $, $\Delta q$ will remain much smaller than
$L$, while for larger values
 of $\eta$ the approximation will break down at a time scale of the
order
of $\gamma^{-1}$.
\par
We should note, that for a large variety of physical systems, it is
not
necessary to assume weak nonlinearities,
in order to have a large value of $L$. Systems with potentials
corresponding
to a spatial average of many microscopic
 degrees of freedom have typical values for $L$ that can be said to correspond
to
a macroscopic scale.
\par
A relevant question is to give an estimation of the error term $ C
\hbar^{\lambda}$
of our approximation.  For the case
 of the closed system, a very rough estimation is
$$ \hbar^{\lambda} \int_0^t dt |V^{(3)}(q(t))| \Delta \dot{q}(t)  $$
in dimensionless units \cite{Omn2}. When considering the open system
case we can see
from the detailed study of the
 proof,  that a term of $Tr \rho^2 < 1$ enters the right hand side of
the inequalities,
thus rendering our
 approximation better. Also for U-shaped potential the effect of the
dissipation is to
make the values of the third
 derivatives of the potential smaller in absolute value. For the time,
that the
increase in the uncertainty due to the
 diffusion is small,
 we expect our approximation to be better when taking into account the
coupling
to the environment.
 On the other hand, at times larger than the typical time where
thermal
fluctuations overcome the quantum ones,
 the more effective spread of the density matrix makes the effect of
non-linearities
more important, and thus
reduces the accuracy of the gaussian approximation. Therefore, we arrive
at the following picture:
\par
 At times less than $(\hbar/ \gamma k T)^{\frac{1}{2}}$ the gaussian
approximation
is better when the
environment is
taken into account.
 \par
 At larger times because of the effects of diffusion, the
approximation becomes
gradually worse, since diffusion
begins  to affect strongly the uncertainty in the position.
\par
Finally we should make a remark on a problem that may arise when
having a closer
look at the proof of the theorem.
We know that the master equation (2.11) does not preserve positivity at
 very short time scales \cite{AnHa,Amb}. This is a result of taking
the cut-off
$\Lambda$ to infinity.
Our proof, being based on a discretization of the time and taking the
continuous
limit in the end, might be inadequate for
  a class of initial states. We can avoid this problem, by adding a
term
$ \eta [p,[p,\rho]]$ in the master equation
(like the one appearing in \cite{Dek}). The master equation thus
becomes
positive and this term does not change the
nature of the proof. In the end we can set $\eta$ equal to zero, an
approximation that is valid in the high temperature regime.
\pagebreak

\section {Evolution of quasiprojectors}
  \par
In this section we study the time evolution of the quasiprojectors
$P$ introduced in Section 3. We shall show that as long as the
corresponding phase space cells remain large and regular, they
evolve according to the classical equations of motion with
dissipation. One expects that the coupling to an environment will
induce
a degradation in the
quasiprojectors as they evolve. This is reflected in an increase of
the size of the margin, which implies a loss of
predictability.
\par
We  work in the equivalent of the Heisenberg picture for open
systems.
That is, we assume that the density matrix
is time independent and that the effect of the evolution is contained
in the operators.
Let us further assume that in this picture a (bounded) operator $P$
evolves under the action of an one parameter
 semigroup with generator ${\cal M}$. The simple correspondence between
${\cal M}$ and ${\cal L}$ that exists in
unitary evolution is lost here.
\par
We determine ${\cal M}$ by demanding that the probabilities are the
same
in both pictures. This means :
\begin{equation}
 Tr( e^{{\cal M}t}[P]  \rho ) = Tr(P  e^{{\cal L}t}[\rho])
\end{equation}
which is equivalent to demanding :
\begin{equation}
 Tr( {\cal M}[P] \rho) = Tr ( P {\cal L}[\rho])
\end{equation}
Having the expression (2.11) for ${\cal L}$ it is easy to show that :
\begin{eqnarray}
 \frac{\partial P}{\partial t} = {\cal M}[P] = &-& \frac{1}{i \hbar}
[\frac{p^2}{2 M} + V(x), P]   \nonumber \\
 &-& \frac{\gamma}{i \hbar} \{p,[P,x]\} - \frac{D}{\hbar^2} [x,[x,P]]
\end{eqnarray}

 The relation of this equation to the one for $\rho$ is more
clarifying in the position representation :
\begin{eqnarray}
 \frac{\partial}{\partial t} P(x,y) = \frac{\hbar}{2 M i} (
\frac{\partial^2}
{\partial x^2} - \frac{\partial^2}{\partial y^2}) P(x,y)
 + \gamma (x-y) (\frac{\partial}{\partial x}  - \frac{\partial}
{\partial y}) P(x,y) \nonumber \\
 - \frac{1}{i \hbar} [V(x) - V(y)] P(x,y) - 2 \gamma P(x,y) -
\frac{D}{\hbar^2} (x-y)^2 P(x,y)
\end{eqnarray}

\par
Comparing this equation to the one for the $\rho$, we notice the
change of sign
in the terms corresponding to the dynamical evolution and the
appearance
of the extra term $ - 2 \gamma P$.
 Because of this term the equation fails to preserve the trace. This is
an expected feature, when studying the evolution
 of quasiprojectors, since classical dissipative evolution does not
preserve the phase space area.
\par
We can study now the evolution of gaussian operators $W(\Sigma,F,r,q,p)$ :
\begin{eqnarray}
 <x|W|y> = (\frac{\Sigma}{2 \pi \hbar})^{\frac{1}{2}} \quad
\exp [ -\frac{\Sigma}{2 \hbar} (\frac{x+y}{2}-q)^2  \nonumber \\
 - \frac{F}{2 \hbar} (x - y)^2 - \frac{i r \Sigma}{2 \hbar}
( \frac{x+y}{2} - q) (x-y) + \frac{ip}{\hbar} (x-y) ]
\end{eqnarray}
It is straightforward to show, using the treatment found in the appendix,
that within an error of $C(\hbar/LP)^{\lambda}$ in the Hilbert-Schmidt norm,
\begin{equation}
  e^{{\cal M}t}[W(\Sigma_0,F_0,r_0,q_0,p_0)] = J(t) \
W(\Sigma(t),F(t),r(t),q(t),p(t))
\end{equation}
where $\Sigma(t)$, $r(t)$, $F(t)$, $q(t)$, $p(t)$ are the solutions to the
equations,
\begin{eqnarray}
 \dot{q} &=& - \frac{p}{M} \\
 \dot{p} &=& 2 \gamma p + V'(q) \\
 \dot{\Sigma} &=& - \frac{1}{M} \Sigma^2 r  \\
 \dot{F} &=& - \frac{1}{M} \Sigma F r + 4 \gamma F + \frac{2 D}{\hbar}  \\
 \dot{r} &=& \frac{\Sigma r^2}{2 M} + \frac{2}{M} F + 2 \gamma r -
\frac{2}
{\Sigma} V^{(2)}(q)
\end{eqnarray}
under the conditions
\begin{equation}
 ( \Sigma(0),F(0),r(0),q(0),p(0) ) = ( \Sigma_0,F_0,r_0,q_0,p_0)
\end{equation}
The equations for $q$ and $p$ are the backwards classical equations of

motion (not the time reverse) and
$J(t)$ is the Jacobian of the transformation from $(q_0,p_0)$ to $(q(t),p(t))$
\begin{equation}
 J(t) = \frac{\partial (q(t),p(t))}{\partial (q_0,p_0)} = e^{2 \gamma t}
\end{equation}
\par
Now let us consider the evolution of a quasiprojector $P_{\Gamma}$, associated
with a phase  space cell $\Gamma$
which is regular to order $\epsilon$. Consider an operator that at $t = 0$ has
maximum resolution in phase space. We thus have at $ t = 0 $
\begin{equation}
 P_{\Gamma} = \int_{\Gamma} \frac{dp_0 dq_0}{2 \pi \hbar}
\quad W(\Sigma_0,\Sigma_0/4,r_0,q_0,p_0)
\end{equation}
As it evolves under the action of $e^{{\cal M}t}$ this projector becomes
\begin{equation}
 e^{{\cal M}t}[P_{\Gamma}] = J(t) \int_{\Gamma} \frac{dp_0 dq_0}
{2 \pi \hbar} \quad W(\Sigma(t),F(t),r(t),q_{cl}(-t),p_{cl}(-t))
\end{equation}
within an error of order
$$(\hbar/LP)^{\lambda}  Tr P_{\Gamma}  $$
Performing a transformation from $(q_0,p_0)$ to
$(q_{cl}(-t),p_{cl}(-t))$
we readily verify that :
\begin{equation}
 e^{{\cal M}t}[P_{\Gamma}] = \int_{\Gamma_t} \frac{dq dp}{2 \pi \hbar}
W(\Sigma(t),F(t),r(t),q,p) = P'_{\Gamma_t} + O(\epsilon')
\end{equation}
where $\Gamma_t$ is the phase space cell obtained from $\Gamma$ through
the classical equations of motion and $\epsilon' = ({\cal A}/LP)^{\frac{1}{2}}$
determines the degree of regularity of $\Gamma_t$. Here we have:
\begin{equation}
 {\cal A} = \frac{\hbar}{2} \\ \frac{4 F(t)}{\Sigma(t)}
\end{equation}
is the Wigner function area. $P'$ is clearly another quasiprojector

associated with the phase space cell $\Gamma_t$.
Essentially we have :
\begin{equation}
\| e^{{\cal M}t}[P_{\Gamma}] - P'_{\Gamma_t} \|_{HS} < c
(\hbar/LP)^{\frac{1}{2}}
TrP_{\Gamma}
\end{equation}
 From (3.4) we know that within an error of $\epsilon'$, $P'_{\Gamma}$ is
equal to $P_{\Gamma_t}$, where $P_{\Gamma_t}$ is the quasiprojector
defined
through the gaussian operator with
parameters $(\Sigma_0,4 \Sigma_0,r_0)$
\begin{equation}
 \| P'_{\Gamma_t} - P_{\Gamma_t} \|_{HS} \leq Tr | P'_{\Gamma_t} -
P_{\Gamma_t}|
< c'  \epsilon'  Tr P_{\Gamma_t}
\end{equation}
\par
Assuming that during the evolution the error of the gaussian
approximation is
less than $\epsilon'$, we conclude from
 (5.18) and (5.19) that
\begin{equation}
 \|e^{{\cal M}t}[P_{\Gamma}] - P_{\Gamma_t}  \|_{HS} < C \epsilon' TrP_{\Gamma}
\end{equation}
\par
This is our main result: A projector $P_{\Gamma}$
onto some phase space cell $\Gamma$ evolves under the action of the
dynamical semigroup into a projector $P_{\Gamma_t}$
associated with the phase space cell $\Gamma_t$, which is obtained
from $\Gamma$ through the classical dissipative equations of motion.
This is an approximate result, accurate within an error of order
$\epsilon'$. This parameter increases with time solely because
of the noise induced by the environment. This implies that the
margin of the quasiprojector increases with time evolution and the
predictability becomes gradually worse.
\par
We can say that the effective region in phase space occupied by a
projector $P_{\Gamma}$ consists
 of the cell $\Gamma$ , plus the margin. Thus  the effective region
the projector occupies increases in a way that has
nothing to do with the corresponding classical equations of motion
( unlike the shrinking due to dissipation which is a
classical feature). A measure for this increase, which can be
attributed
solely to the diffusion, is
\begin{equation}
\mu = \frac{[\Gamma_t]+[M']}{[\Gamma_t] + [M]} = \frac{1 +
\epsilon'}{1
+ \epsilon} \simeq
1 + \epsilon' - \epsilon \simeq 1 + (\frac{{\cal
A}}{LP})^{\frac{1}{2}}
( 1 - (\hbar/{\cal A})^{\frac{1}{2}})
\end{equation}
\par We can view this result as follows:
Sufficiently coarse grained observables (projectors onto large phase
space cells)
evolve under
classical deterministic equations plus the action of noise terms.
These terms, stemming from the ignored degrees
 of freedom of both the system and the environment induce an error of
order
$\epsilon'$ to the classical equations
 of motions. The parameter measuring the size of the noise terms and
correspondingly the loss of predictability
 is the Wigner function area ${\cal A}$ \cite{AnHa}, which in general
is an increasing function of time.
\par
It may be shown that in the short time limit ($t << \gamma^{-1}$) for
all initial operators (5.14) it is given by  \cite{AnHa}   :
\begin{equation}
 {\cal A}^2(t) = \frac{\hbar^2}{4} + \frac{32}{3} \frac{\gamma^2 k^2
T^2 }
{\hbar^2} t^4
\end{equation}
\par
We obtain the same result, when considering the backward time
evolution of
the quasiprojectors.

\section{Histories}
\subsection{Construction of the decoherence functional}
\par
Using our previous results we now construct the phase space histories for
the Brownian particle and study their decoherence properties.
\par
We have found how the quasiprojectors associated with phase space cells evolve
in
time. We can use this knowledge in the calculation of the decoherence
functional. Our results are of use only when it can be written as:
\begin{eqnarray}
 D(\alpha,\alpha') = Tr_{s}[P^n_{\alpha_n} K_{t_{n-1}}^{t_n}
[ \ldots P^1_{\alpha_1} K_{t_0}^{t_1}[\tilde{\rho}]
P^1_{\alpha'_1} \ldots] P^n_{\alpha'_n} ]
\end{eqnarray}
in terms of a superoperator $K_t^{t'}$ in the space of the reduced density
matrices. This is actually valid in the Markovian regime \cite{PaZu}.
\par
In the Caldeira -Leggett model the environment is taken to be
essentially
infinite and thus the environmental degrees
of freedom remain very close to the state of thermal equilibrium. The
correlations created will in general be of
 the order of $\Lambda^{-1}$ and since the evolution has no memory
their
effect will be negligible in a time scale
 of order $\Lambda^{-1}$. Remembering that in our regime we have
assumed
$ k T >> \hbar \Lambda$, we conclude
that if the time interval between the two projections satisfies :
$$ t_2 - t_1 > O(\hbar/kT) $$
we can write the decoherence functional for two-time histories as
\begin{equation}
 D(\alpha,\alpha') = \delta_{\alpha_2 \alpha_2'} Tr ( P^2_{\alpha_2}
e^{{\cal L}(t_2-t_1)} [ P^1_{\alpha_1}] \tilde{\rho}_{t_1}
 P^1_{\alpha_1'})
\end{equation}
or using the definition of the generator ${\cal M}$
\begin{equation}
 D(\alpha,\alpha') = \delta_{\alpha_2 \alpha_2'} Tr ( e^{-{\cal M}
(t_2-t_1)}[P^2_{\alpha_2}] P^1_{\alpha_1} \tilde{\rho}_{t_1}
 P^1_{\alpha_1'})
\end{equation}
We can assign a probability measure
\begin{equation}
 p(C_{\alpha}) = Tr(( e^{-{\cal M}(t_2-t_1)}[P^2_{\alpha_2}]
P^1_{\alpha_1}
\tilde{\rho}_{t_1}
P^1_{\alpha_1})
\end{equation}
to the history $C_{\alpha}$ if the probability sum rules are
satisfied.
A sufficient condition for this is the
 vanishing
of the off-diagonal elements of the decoherence functional.
\subsection{Approximate decoherence and determinism}
\par
The sum rules are not satisfied exactly in our case, but only within an error
of order
   $\epsilon$. A good condition for this approximate consistency is :
\begin{eqnarray}
 | Tr  ( e^{-{\cal M}(t_2-t_1)}[P^2_{\alpha_2}] P^1_{\alpha_1}
\tilde{\rho}_{t_1}
P^1_{\alpha_1'}) | <  \nonumber \\
\epsilon \quad \sum_{\alpha_1}  Tr  ( e^{-{\cal
M}(t_2-t_1)}[P^2_{\alpha_2}]
P^1_{\alpha_1} \tilde{\rho}_{t_1} P^1_{\alpha_1})
\end{eqnarray}
\par
Let us concentrate on the particular case of histories that correspond

to the evolution of phase space cells along
the classical trajectories. That means, we consider the set of histories
$C_1 = ( P_1,P_2 )$ and $C_2 = (\bar{P}_1,P_2) $
where $P_1$ and $P_2$ are quasiprojectors associated with the cells
$\Gamma_1$
and $\Gamma_2$ and $\Gamma_2$
 is the cell obtained from the classical evolution of $\Gamma_1$. In
order
to achieve minimum error in our
 estimations we restrict our quasiprojectors to have minimum resolution in
phase space.
The condition (6.8) reads :
\begin{eqnarray}
  Tr (e^{-{\cal M}(t_2-t_1)}[ P_2] P_1 \tilde{\rho}_{t_1} \bar{P}_1)
<  \hspace{6cm} \nonumber \\
 \epsilon ( Tr( (e^{-{\cal M}(t_2-t_1)}[ P_2] P_1 \tilde{\rho}_{t_1} P_1)
 +  Tr (e^{-{\cal M}(t_2-t_1)}[P_2] \bar{P}_1 \tilde{\rho}_{t_1} \bar{P}_1) )
\end{eqnarray}

{}From the analysis in the previous section, we know that within an error of
order $({\cal A}(t_2-t_1)/LP)^{\frac{1}{2}} Tr P_1$
\begin{equation}
e^{-{\cal M}(t_2-t_1)}[ P_2] = P_1
\end{equation}
This means that the left hand side in (6.6) is of order
$$({\cal A}(t_2-t_1)/LP)^{\frac{1}{2}} Tr P_1$$ and the right hand side of
order
$$\epsilon  Tr  (P_1 \tilde{\rho}_{t_1}) < \epsilon  TrP_1$$
Thus the probability sum rules are satisfied within an order
 $$ \epsilon = ({\cal A}(t_2-t_1)/LP)^{\frac{1}{2}} $$
${\cal A}$ takes values between $\hbar/2$ and its asymptotic value
which
is of the order of the thermal fluctuations in phase
space. Therefore, for a macroscopic phase space cell $\epsilon$ is a
small number and decoherence is good. It is clear that
the degree of decoherence is better in the case of a hamiltonian system.
We can thus compute the conditional probability, that the particle
being within
$\Gamma_1$ at time $t_1$ will be
within $\Gamma_2$ at time $t_2$. This is clearly,
\begin{equation}
 p(\Gamma_1,t_1 \rightarrow \Gamma_2 , t_2) = 1 - O(\epsilon)
\end{equation}
Therefore up to an order of $\epsilon$ there is good agreement between the
predictions of classical and quantum dynamics for the evolution of the Brownian
particle. This agreement is best when the phase space cells are regular and the
time
evolution preserves this regularity. These results are independent
of the initial state of the system. One might expect that the degree
of decoherence
would be in inverse relation to the degree of
predictability. Here, they are essentially the same, as in the case
where the
environment is not present.
\par
The extension of this result to n-time histories is
straightforward. Consider
the history
$C = ( P_1,t_1;P_2,t_2 ;\ldots ; P_n,t_n)$ where $ P_i$ is the maximum

resolution quasiprojector associated
 with a cell $\Gamma_i$ and $\Gamma_{i+1}$ is obtained from $\Gamma_i$

through the classical equations
 of motion. We construct from C a set of histories by replacing one or

more of the $P_i$ 's with $\bar{P}_i$,
and define the parameter $\epsilon = \max ({\cal A}(t_i -
t_{i-1})/LP)^{\frac{1}{2}}$.
If we assume  that for the intermeddiate
 times
  $$\min (t_{i+1} - t_i) > O(\hbar/kT)$$
 and that the dynamics preserve the regularity of the cells to an order
 $\epsilon$, then this set of histories satisfies the probability
summation rules to this order.

\section{Conclusions}
\par
In this paper, we have studied histories corresponding to the
evolution o
f phase space cells in quantum Brownian
 motion models with ohmic dissipation and Markovian dynamics.
We constructed these histories using the
generalization (3.14) of the quasiprojectors used by Omn\`es.
The estimation (4.1) for the validity of the gaussian approximation
has enabled us to establish, how those
 quasiprojectors evolve.
\par
We showed that as long as the phase space cell is large and regular,
histories corresponding to the evolution
 according to the classical deterministic equations of motion approximately
satisfy
the sum rules. The order of
magnitude of the corresponding error depends only on the Wigner
function
area ${\cal A}$ (a function only of the time
difference between projections). ${\cal A}(t)$ can be computed as the
solution of the system (5.7 - 5.11) under the initial
condition (5.12). It starts as containing purely quantum uncertainties

at $t = 0$ and at a timescale of $\gamma^{-1}$ becomes
of the order of magnitude of its asymptotic value, which corresponds
to
thermal fluctuations.
\par
There are three timescales in our model, which determine qualitative
changes in the description of the dynamics:
the Markovian time $t_M = \hbar/kT$, the decoherence time
$t_d = (\hbar/\gamma kT)^{\frac{1}{2}}$ and the relaxation time
$t_r = \gamma^{-1}$. For times $t < < t_M$ dynamics are not Markovian
and therefore we cannot talk about predictability. For
times $t$ such that $t_M < t < t_d $, sufficiently large phase space
cells evolve almost under deterministic equations of motion
and the details of the potential are not of importance in the increase

of the fluctuations.
At $t = t_d$ the thermal fluctuations overcome the quantum
ones \cite{AnHa}. For $t > t_d$ the degree of decoherence and
predictability gradually worsens and is destroyed for cells with area

less than the thermal uncertainty. For these cells any
sense of predictability has been lost at $t = t_r$. In the contrary
sufficiently macroscopic cells, continue to evolve within a
good approximation according to the classical equations of motion
even at times larger than $\gamma^{-1}$. Eventually at
long times, approximately deterministic evolution will break down,
because the dissipative nature of the evolution tends to
shrink the phase space cells, until their area is of the order of
magnitude of the thermal fluctuations. We therefore see that the
breakdown of predictability is dependent mainly on the parameters
of the environment and the structure of the potential is
largely irrelevant, unlike the unitary case. This comes probably
from the fact that we have considered an one dimensional
problem. It is to be expected that in a system with more degrees
of freedom, time parameters of the potential (i.e the Lyapunoff
exponents of the classical solutions) will play a more important
role in the determination of when classical predictability
breaks down.
\par
Perhaps contrary to our expectations, the coupling to the environment
tends to make the decoherence properties of histories
 worse. This means that histories corresponding to evolution of phase
space cells are not the ones that give the
 sharpest correspondence with a deterministic classical description.
This is to be expected, since the classical
 description of  quantum brownian motion is that of a stochastic
process in phase space.
\par
Essentially, there are two mechanisms that produce decoherence.
One is the interaction with an environment and the other
is the existence of phase space projection operators that have
an intrinsic almost deterministic time evolution. In the case of
purely hamiltonian
dynamics only the latter
appears. In our case, we have both, but it is clear that again the
latter is the dominant one. This is attributed to the simplicity of
the corresponding classical equations, in which the effect of the
environment is contained only within the term $- 2 \gamma p$.
It is therefore natural that observables close to the ones used
in purely Hamiltonian systems will decohere giving rise to
approximate predictability.
\par
With those results we have achieved three things: First, we have
generalized the results of Omn\`es on classical determinism for a
class of open systems. In fact, we can generalize our
results are valid for any Markovian equation of the Lindblad type
\cite{Lin}, with environment operators $L_n$ linear in position
and momentum. Second, we have verified the validity of classical
equations of motion with linear dissipation, as a
consequance of the underlying quantum mechanical evolution and
sufficient coarse-graining. Third, taking quantum Brownian
motion as a toy model, we have obtained a picture of the
behaviour of collective variables in many body systems and the
quasiclassical behaviour of these variables.
\par
Finally, we would like to compare our results with the ones obtained
in \cite{HaZo} using the quantum state diffusion
picture  for the quantum Brownian motion. For the case of  two-time
phase space histories
$C = (\Gamma_1,\Gamma_2)$ and under the assumption that they decohere,
the authors gave for the associated probability the expression :
\begin{eqnarray}
p(C) = \int_{\Gamma_2} dp_2 dq_2 \int_{\Gamma_1} dp_1 dq_1
J(p_2,q_2,t_2|p_1,q_1,t_1) f(p_1,q_1,t_1)
\end{eqnarray}
where $f(p,q,t) $ is a classical probability distribution satisfying
the Fokker-Planck equation and $J$ is the associated
 propagator. We can verify that this result is equivalent to ours
in the particular case of an harmonic potential,
 where the propagator is gaussian centered around the classical path.
If the volume of the phase space cells
 is much larger than the Wigner function area ${\cal A}(t)$ associated

with the propagator \cite{AnHa}, then the propagation is
 within an error of order $\epsilon = ({\cal A}/LP)^{\frac{1}{2}}$ a
$\delta$-function around the classical path and
the conditional probability is found
$$p(\Gamma_1,t_1 \rightarrow \Gamma_2,t_2) = 1 - O(\epsilon)$$

\section{Aknowledgements}
I would like to thank S.\ Schreckenberg for a useful suggestion and in
particular
J.\ J.\ Halliwell for suggesting this project and for many discussions and
encouragement during the research.
\par
This research was supported by the Greek State Scholarship Foundation.

\begin{appendix}
\section{Proof of the theorem}
\par
We are going to give the proof for the theorem stated in section 4.
It follows the reasoning of the similar proof of Hagedorn for the case

of closed system, where the reader can refer for further details.
\par
The evolution equation for the density matrices in the Markowian
regime  (2.11)
corresponds
to the action of an one-parameter semigroup with generator ${\cal L}$
on the state space .
We can write :
$$ {\cal L} = {\cal S} + {\cal U} $$
where ${\cal S}$ and ${\cal U}$ are generators of one parameter semigroups
given by :
\begin{equation}
{\cal S}[\rho] = \frac{1}{i \hbar} [p^2/2M,\rho] + \frac{\gamma}{ i
\hbar}
[x,\{\rho,p\}] - \frac{D}{\hbar^2} [x,[x,\rho]]
\end{equation}

\begin{equation}
 {\cal U}[\rho] = \frac{1}{i \hbar} [V(x),\rho]
\end{equation}

Splitting ${\cal L}$ into ${\cal S}$ and ${\cal U}$ will enable us to
use
the Trotter product formula to determine
$e^{{\cal L}t}$ in terms of $e^{{\cal S}t}$ and $e^{{\cal U}t}$.
\par
We want to establish, how the class of states described by the
gaussian
density matrices
(3.11) evolve under ${\cal S}$ and ${\cal U}$.  \\
{\bf 1. Evolution under S}
\par
Let us consider the solution of :
$$ \frac{\partial \rho}{\partial t} = {\cal S}[\rho] $$
with initial conditions $$\rho = \rho(\Sigma_0,F_0,r_0,q_0,p_0) $$ at
time $ t = 0 $.
This is just the solution
of the master equation for the free particle. For short times t
($t < \tau << \gamma^{-1}$ ) we have that :
\begin{equation}
\| e^{{\cal S}t} [ \rho(\Sigma_0,F_0,r_0,q_0,p_0)] -
\rho(\Sigma,F,r,q,p) \|_{H.S.}  < K (t/\tau)^2
\end{equation}
where :
\begin{eqnarray}
 q &=& q_0 + \frac{p_0}{M} t     \\
 p &=& p_0 - 2 \gamma p_0 t     \\
 \Sigma &=& \Sigma_0 + ( \frac{2}{M} \Sigma_0^2 r_0 ) t   \\
 F &=& F_0 + ( \frac{1}{M} \Sigma_0 F_0 r_0 - 4 \gamma F_0+
\frac{2 D}{\hbar} ) t    \\
 r &=& r_0 - (\frac{1}{2 M} \Sigma_0 r_0^2 -\frac{2 \gamma}{M} F_0
- 2 \gamma r_0 ) t
\end{eqnarray}
\\
{\bf 2. Evolution under U - quadratic potential}
\par
Consider  the case of the most general  quadratic potential :
\begin{equation}
 V(x) = a (x-q) + \frac{1}{2} (x-q)^2
\end{equation}
We can easily verify :
\begin{equation}
 e^{{\cal U}t} [ \rho(\Sigma_0,F_0,r_0,q_0,p_0) ] =
\rho( \Sigma_0,F_0,r_0 + b t,q_0,p_0 - a t)
\end{equation}
\\
{\bf 3. Evolution under U - general potential}

For the case of a general potential satisfying the conditions we
stated in section 4, we define :
\begin{equation}
 W_q(x) = V(q) + V'(q) + \frac{1}{2} V^{(2)}(q) (x - q)^2
\end{equation}

Denote by ${\cal U}_V$ and ${\cal U}_{W_q}$ the generators
corresponding
to $V(x)$ and $W_q(x)$ respectively.
We have :
\begin{equation}
 \| e^{{\cal U}_V  t}[\rho] - e^{{\cal U}_{W_q} t} [ \rho] \|^2_{H.S}
\leq \frac{t^2}{\hbar^2}
(\rho,({\cal U}_V-{\cal U}_{W_q})^2[\rho])
\end{equation}
in the Hilbert-Schmidt inner product.
Writing : $ {\cal O} = | U_V - U_{W_q} |   $
we see that the right hand side reads :
\begin{equation}
 \frac{t^2}{\hbar^2} ( {\cal O}[\rho],{\cal O}[\rho]) = \frac{ 2
t^2}{\hbar^2}
\quad Tr(\rho^2 O^2- \rho O \rho O )
 \end{equation}
where $O$ is the operator $ |V(x) - W_q(x)| $ acting on the Hilbert
space
$H_{\psi}$.
In the coordinate representation this quantity reads :
\begin{eqnarray}
 \frac{2 t^2}{\hbar^2} \int dx dy \rho(x,y) \rho(y,x) O(x) [O(x)-O(y)]
\leq
\hspace{4cm}\nonumber \\
 \frac{2 t^2}{\hbar^2} \int dx \langle x|\rho^2|x \rangle O^2(x) =
\hspace{7.5cm}  \nonumber \\
 \frac{2 t^2}{\hbar^2} \frac{\Sigma}{2 \pi \hbar}
(\frac{\pi \hbar}{\Sigma/4 + F})^{\frac{1}{2}} \int dx \exp
  [ -\frac{1}{\hbar}\frac{\Sigma F}{\Sigma/4 + F} (x - q)^2 ] O^2(x)
\hspace{2cm}
\end{eqnarray}

The hypothesis that $V(x)$ s uniformly Lifschitz implies that for
$q \epsilon K$ (some compact subset of ${\cal R}$)
and $|x-y|<1$ there exists $\beta$ such that:
$$ | V(x) - W_q(x)| < \beta |x - y|^3  $$
Following the treatment of Hagedorn we can split the integral into two sectors:
$ D = \{ x ; |x-q| \leq \hbar^{\alpha}  \} $ and $ \bar{D} = {\cal R} - D $
where $ \alpha < \frac{1}{2}$.
\par
The first part reads:
\begin{eqnarray}
 \frac{2 t^2}{\hbar^2} \frac{\Sigma}{2 \pi \hbar}
(\frac{\pi \hbar}{\Sigma/4 + F})^{\frac{1}{2}} \int_D dx
\exp [ - \frac{1}{\hbar} \frac{\Sigma F}{\Sigma/4 + F} (x-q)^2 ]
O^2(x)
\leq  \nonumber \\
 \frac{2 t^2}{\hbar^2} \beta^2 \frac{\Sigma}{2 \pi \hbar}
(\frac{\pi \hbar}{\Sigma/4 + F})^{\frac{1}{2}}
 \int_D dx    \exp [ - \frac{1}{\hbar} \frac{\Sigma F}{\Sigma/4 + F}
(x-q)^2 ] | x - q|^6
\nonumber \\
  \leq
 \frac{2 t^2}{\hbar^2} \beta^2 (Tr \rho_0^2)^{\frac{1}{2}} \hbar^{6 \alpha}
\end{eqnarray}
The second part gives :
\begin{eqnarray}
 \frac{2 t^2}{\hbar^2} \frac{\Sigma}{2 \pi \hbar}
(\frac{\pi \hbar}{\Sigma/4 + F})^{\frac{1}{2}} \int_{\bar{D}} dx
\exp [ - \frac{1}{\hbar} \frac{\Sigma F}{\Sigma/4 + F} (x-q)^2 ]
O^2(x)
\leq  \nonumber \\
  \frac{2 t^2}{\hbar^2} C_1^2  \frac{\Sigma}{2 \pi \hbar}
(\frac{\pi \hbar}{\Sigma/4 + F})^{\frac{1}{2}}
 \int_{\bar{D}} dx \exp [ - \frac{1}{\hbar} \frac{\Sigma F}{\Sigma/4 +
F} (x-q)^2 ]
e^{2 M x^2}  \leq \nonumber \\
 \frac{2 t^2}{\hbar^2} C_1^2  \frac{\Sigma}{2 \pi \hbar} (\frac{\pi
\hbar}
{\Sigma/4 + F})^{\frac{1}{2}}
\int_{\bar{D}} dx \exp \{[ - \frac{1}{2 \hbar} \frac{\Sigma
F}{\Sigma/4 + F}
+ 2 M ](x-q)^2 \}
\nonumber \\
  \int_{\bar{D}} dx
\exp [ - \frac{1}{2 \hbar} \frac{\Sigma F}{\Sigma/4 + F} (x-q)^2 ]
\end{eqnarray}
where in the last step we used the Cauchy-Schwarz inequality at
$L^2(\bar{D})$ (the space of real-valued square integrable functions
on $\bar{D}$).
This quantity is smaller than
\begin{equation}
 \frac{2  t^2}{\hbar^2}  (2 Tr \rho^2)^{\frac{1}{2}}
\exp [ M \hbar^{ 2 \alpha} - \hbar^{2 \alpha -1} ]
\end{equation}
Assume that
\begin{equation}
(\frac{\Sigma F}{\Sigma + 4 F})^{-1} \leq N \hbar^{-2p}
\end{equation}
for some $N > 0$ and $p \leq \alpha$. This implies the existence of
$ \delta > 0 $ such that:
$$ M - (\frac{\Sigma F}{\Sigma + 4 F}) \hbar^{-1} \leq M - \hbar^{2 p -1}/N^2
$$
for $ \hbar < \delta  $ and taking $ \alpha < \frac{1}{2} - p $
there exists constant $c_2$ such that :
$$  \exp [ M \hbar^{2 \alpha} - \hbar^{2p+ 2 \alpha - 1} / N^2 ]
< c_2 \hbar^{6 \alpha}  $$
Thus we establish that when $ (\frac{\Sigma F}{\Sigma + 4 F})^{-1}
\leq N \hbar^{-2p}$, $q \epsilon K$, for each $N>0$
, $p<\frac{1}{2}$, $\alpha < \frac{1}{2} - p$ and $ t > 0$ there
exists
$C>0$ and $\delta > 0  $ , such that for
 $\hbar < \delta$ :
\begin{equation}
\| e^{{\cal U}_V t}[\rho] - e^{{\cal U}_{W_q} t} [ \rho] \|^2_{H.S}
< C \hbar^{3 \alpha - 1} (Tr \rho^2)^{\frac{1}{2}}  t
\end{equation}
\par
We now have expressions for the evolution induced by the generators
${\cal S}$ and ${\cal U}$. We will combine them by using
the Trotter product formula.
\\
{\bf 4. Conclusion of the proof}
\par
The hypotheses on $V(x)$ ensure the existence of a bounded solution to
the equations (4.2-4.6) for $ t < T $. We can rewrite our previous result as :
\begin{equation}
\| e^{{\cal U}_Vt}[\rho_0] - e^{{\cal U}_{W_q}t} [ \rho_0] \|^2_{HS}
< C' \hbar^{\lambda}/3 T
\end{equation}
where $ \lambda = 3 \alpha - 1 < \frac{1}{2}  $.
We discretize the system (4.2-4.7) taking time step $\delta t =t/N$ .
We have :
\begin{eqnarray}
 q_N(n) &=& q_0 + \sum_{i=1}^{n} \frac{p_N(i)}{M} \delta t  \\
 p_N(n) &=& p_0 - \sum_{i=1}^{n} [ V'(q_N(i)) + 2 \gamma p_N(i) ] \delta t  \\
 \Sigma_N(n) &=& \Sigma_0 + \frac{1}{M} \sum_{i=1}^{n} [\Sigma_N^2(i)
r_N(i)]
\delta t  \\
 F_N(n) &=& F_0 + \sum_{i=1}^{n} [\frac{1}{M} \Sigma_N(i) F_N(i) r_n(i) -
4 \gamma F_N(i) + \frac{ 2 D}{\hbar}] \delta t \\
 r_N(n) &=& r_0 - \sum_{i=1}^{n} [ \frac{ \Sigma_N(i) r_N^2(i)}{2 M}
+ \frac{2 F_N(i)}{M}
+ 2 \gamma r_N(i) \nonumber \\
&-& \frac{2}{\Sigma_N(i)} V^{(2)}(q_N(i))] \delta t  \\
\end{eqnarray}
\par
Due to uniform convergence, we can always find $N_1$ such that for $N > N_1$:
\begin{eqnarray}
 \| \rho(\Sigma_N(N),F_N(N),r_N(N),q_N(N),p_N(N)) \nonumber \\
- \rho(\Sigma(t),F(t),r(t),q(t),p(t)) \|_{HS} < C' \hbar^{\lambda}/3
\end{eqnarray}
\par
The Trotter product formula ensures the existence of $N_2$ such that for $N >
N_2$ :
\begin{equation}
 \| (e^{{\cal L}t} - [e^{{\cal S}t/N} e^{{\cal U}t/N}]^N)[\rho_0]
\|_{HS}
< C' \hbar^{\lambda}/3
\end{equation}
For $ N > max(N_1,N_2) $ we have :
\begin{equation}
 \| [ e^{{\cal S}t/N} e^{{\cal U}t/N} - e^{{\cal S}t/N} e^{{\cal
W}_{q_0}t/N} ]
[\rho(\Sigma_0,F_0,r_0,q_0,p_0)] \|_{HS} < C' \hbar^{\lambda}/3
\end{equation}
We also have :
\begin{eqnarray}
\| e^{{\cal S}t/N} e^{{\cal U}t/N} [\rho(\Sigma_0,F_0,r_0,q_0,p_0)] -
\rho(\Sigma_N(1),F_N(1),r_N(1),q_N(1),p_N(1)) \|_{HS}   \nonumber \\
 \leq \| [e^{{\cal S}t/N} e^{{\cal U}t/N} -e^{{\cal S}t/N} e^{{\cal
W}_{q_0}t/N} ]
[\rho(\Sigma_0,F_0,r_0,q_0,p_0)] \|_{HS} \hspace{3cm} \nonumber \\
 +\| e^{{\cal S}t/N} \rho(\Sigma_0,F_0,r_0+V^{(2)}(q_0) t/N,q_0,p_0 -
V'(q_0) t/N)
\hspace{3cm}\nonumber \\
-\rho( \Sigma_N(1),F_N(1),r_N(1),q_N(1),p_N(1)) \|_{HS} \hspace{5cm} \nonumber
\\
 \leq C' \hbar^{\lambda} /(3N) + K (t/\tau)^2 N^{-2}  <
C \hbar^{\lambda}/(3N) \hspace{5cm}
\end{eqnarray}
for $ N > N_3 > max(N_1,N_2) $ and with $C > C'$.
\par
For these values of N, we can obtain by iteration :
\begin{eqnarray}
 \| \prod_{i=1}^{n} e^{{\cal S}t/N} e^{{\cal U}t/N}
[\rho(\Sigma_0,F_0,r_0,q_0,p_0)] &-& \rho(\Sigma_N(n),F_N(n),r_N(n),q_N(n),
p_N(n) \|_{HS} \nonumber \\
&\leq& n C \hbar^{\lambda}/(3N)
\end{eqnarray}
Taking the above inequality for $n = N$ and using (9.27) and (9.30)  we arrive
at:
\begin{equation}
\| e^{{\cal L}t} [ \rho(\Sigma_0,F_0,r_0,q_0,p_0)]
- \rho(\Sigma(t),F(t),r(t),q(t),p(t)) \|_{HS} < C \hbar^{\lambda}
\end{equation}

\end{appendix}                         QED
\end{document}